\documentclass[11pt]{article}
\usepackage{amsfonts,amssymb,amsmath}
\usepackage{stmaryrd}
\usepackage[T2A]{fontenc}
\usepackage[cp1251]{inputenc}

\def \TITLE {Renormalization of Green Functions in Configuration Space}

\usepackage{ulem}

\newcommand{\beq}{\begin{equation}}
\newcommand{\eeq}{\end{equation}}
\newcommand{\beqa}{\begin{eqnarray}}
\newcommand{\eeqa}{\end{eqnarray}}
\newcommand{\nn}{\nonumber \\}
\def \podr {&& \hspace{-15pt}}

\def \z {{\mathrm{z}}}
\def \w {{\mathrm{w}}}

\def \x {{\mathrm{x}}}

\def \y {{\mathrm{y}}}

\def \rr {\mbf{\mathrm{r}}}

\def \rR {\mathrm{r}}

\def \sS {\mathrm{s}}

\def \Ss {\mathrsfs{S}}
\def \Dd {\mathrsfs{D}}
\def \R {{\mathbb R}}
\def \C {{\mathbb C}}
\def \Z {{\mathbb Z}}
\def \N {{\mathbb N}}

\def \di {\partial}
\def \mz {\bigr\backslash\hspace{1pt}\{\Mbf{0}\}}

\DeclareMathAlphabet{\mathbbm}{U}{bbm}{m}{n}

\DeclareSymbolFont{ltrs}     {OT1}{pzc}{m}{it}
\DeclareSymbolFont{ltrsa}     {OMS}{cmsy}{m}{n}
\DeclareSymbolFont{ltrsA}{U}{txmia}{m}{it}
\DeclareSymbolFont{symbolsC}{U}{txsyc}{m}{n}
\DeclareSymbolFont{ltrsB}{U}{rsfs}{m}{n}

\DeclareSymbolFontAlphabet{\mfrak}{ltrsA}
\DeclareMathAlphabet{\mathpzc}{OT1}{pzc}{m}{it}
\DeclareMathAlphabet{\mathrsfs}{U}{rsfs}{m}{n}

\def \ID {\text{\it id}}

\def \La {
\left\langle \!\!{\,}^{\mathop{}\limits_{}}_{\mathop{}\limits^{}}\right.}
\def \Ra {
\left. \!\!{\,}^{\mathop{}\limits_{}}_{\mathop{}\limits^{}}\right\rangle}
\newcommand{\vrestr}[2]{\!\left.\raisebox{#1}{$\,$}\!\right|_{\,
\raisebox{1pt}{\small \(#2\)}}}

\newcommand{\mbf}[1]{\ensuremath{\mathchoice
                    {\mbox{\boldmath$\displaystyle\mathbf{\mathit{#1}}$}}
                    {\mbox{\boldmath$\textstyle\mathbf{\mathit{#1}}$}}
                    {\mbox{\boldmath$\scriptstyle\mathbf{\mathit{#1}}$}}
                    {\mbox{\boldmath$\scriptscriptstyle\mathbf{\mathit{#1}}$}}}}

\newcommand{\Mbf}[1]{\ensuremath{\mathchoice
                    {\mbox{\boldmath$\displaystyle\mathbf{#1}$}}
                    {\mbox{\boldmath$\textstyle\mathbf{#1}$}}
                    {\mbox{\boldmath$\scriptstyle\mathbf{#1}$}}
                    {\mbox{\boldmath$\scriptscriptstyle\mathbf{#1}$}}}}

\newcounter{Theorem}
\newcounter{Definition}
\newcounter{Remark}

\def \setcntrs {\setcounter{equation}{0}\setcounter{Theorem}{0}\setcounter{Definition}{0}\setcounter{Remark}{0}}

\renewcommand{\thesection}{\arabic{section}}

\newcommand{\medskp}{

  \medskip

  }

\newenvironment{Theorem}[1][\bf Theorem \arabic{section}.\arabic{Theorem}]{%
        
        \refstepcounter{Theorem}\noindent\textbf{#1.}${}$\hspace{1pt}${}$\it}{}

\newenvironment{Lemma}[1][\bf Lemma \arabic{section}.\arabic{Theorem}]{%
        
        \refstepcounter{Theorem}\noindent\textbf{#1.}${}$\hspace{1pt}${}$\it}{}

\newenvironment{Remark}[1][\bf Remark \arabic{section}.\arabic{Remark}]{%
        
        \refstepcounter{Remark}\noindent\textbf{#1.}${}$\hspace{1pt}${}$}{}

\newcounter{tmpc}
\newlength{\tmplenght}
\newlength{\tmplenghta}
\newlength{\tmplenghtb}
\newlength{\tmplenghtc}

\def\DP{\Dd'}
\def\CI{\mathcal{C}^{\infty}}

\def\scdeg{\text{\rm sc.{\hspace{1pt}}d.}}

\def \Rr {\Mbf{r}}

\def\RMA{R}

\def\SRMA{{\mathop{\RMA}\limits^{\text{\tiny $\bullet$}}}}
\def\Prma{\mathcal{P}}
\def\PRMA{P}

\usepackage{ulem}

\def\PA{{\mathrsfs{O}\hspace{-1pt}}}

\newcommand{\NPR}[1]{:\hspace{-1pt}{#1}\hspace{-2pt}:\hspace{-2pt}}
\def\FSP{\mathcal{F}}
\def\HDP{\widehat{\Dd}\hspace{0.75pt}'\hspace{-3.75pt}}
\def\DOM{\mathfrak{D}}
\def\lvac{\langle 0 |}
\def\rvac{| 0 \rangle}
\def\INFT{}
\def\INFt{}
\def\FSPO{\FSP^{(1)}}
\def\CSL{\mathfrak{C}}
\def\GH{\widehat{G}}
\def\BV{\text{b.v.}}
\newcommand{\VE}[1]{\vec{#1}}
\def\TUBE{\mathcal{T}}
\def\LE{\prec}
\newcommand{\WF}[2]{\text{w.f.}_{#1}\bigl(#2\bigr)}
\def\Cnj{*}
\def\CNj{\star}
\def\CNJ{\hspace{1pt}\raisebox{1pt}{\scriptsize $\CNj$}}

\newcommand{\Section}[2]{%
 \refstepcounter{section}
 \section*{\large \arabic{section}. #1}\label{#2}%
 \addtocontents{toc}{\protect\vspace{-8pt}}
 \addtocontents{toc}{\contentsline {section}{\thesection.\hspace{6pt}{#1}}{\arabic{page}}}\setcntrs}

\def\AFTR{\gtrsim}

\title{Renormalization theory of Feynman amplitudes on configuration spaces}

\author{Nikolay M. Nikolov}

\begin{document}

\maketitle

\thispagestyle{empty}

\vspace{-0.8cm}

\begin{center}
\scriptsize
Institute for Nuclear Research and Nuclear Energy, \\
Tsarigradsko Chaussee 72, BG-1784 Sofia, Bulgaria \\
mitov@inrne.bas.bg
\end{center}

\vspace{0cm}

\begin{abstract}
In a previous paper
``\textit{Anomalies in Quantum Field Theory and Cohomologies of Configuration Spaces}''
(arXiv:0903.0187) we presented a new method for renormalization in
Euclidean configuration spaces based on certain renormalization maps.
This approach
is aimed to serve for
developing an algebraic algorithm
for computing the Gell--Mann--Low  renormalization group action.
In the present work we introduce a modification of the theory of renormalization maps
for the case of Minkowski space and we give the way how it is combined within
the
causal perturbation theory.
\end{abstract}%

\tableofcontents

\Section{Introduction}{Intr}

The causal approach to perturbative quantum field theory (QFT)
originates in the work of Stuckelberg and Bogolubov
and was fully developed by
(and named after) Epstein and Glaser (\cite{EG}, see also \cite{S2}, \cite{DF}).
In this method
the
renormalization is done for the
products of fields (time--ordered, or retarded).
This facilitates the generalization of
the
perturbation theory on manifolds
but still it has the disadvantage of being
rather complicated technically,
especially for concrete calculations.

In paper \cite{N1} (see also its review \cite{N2}) we have found an equivalent construction
to the Epstein--Glaser procedure, which is entirely
set up
in terms of
renormalization of Feynman amplitudes (integrals of functions).
This approach then has the additional advantage
of being independent of concrete models
of quantum fields like
the
$\varphi^4$--theory or
quantum electrodynamics etc.
In this way we get rid of the
technical difficulties present in
a particular
model, in other words, we separate them
from the renormalization problem.
Furthermore, our reformulation of the renormalization problem
makes possible to give a geometric characterization for the
renormalization ambiguity.
Our main goal there was to use this geometric analysis in order to derive
an algebraic algorithm for determining the Gell--Mann--Low  renormalization group action,
i.e., the action of the one parameter group $\R^+$ on the space of coupling constants,
which is induced by the scaling transformations (in terms of formal diffeomorphisms).
In particular, we are interested in algebraic algorithms for calculating the perturbative expansions
of $\beta$--functions and anomalous dimensions.

In the present paper we introduce a Minkowski space version
of the theory of renormalization maps developed in \cite{N1}
and we also combine this theory
with
the
causal perturbation theory.

\Section{Axiomatic properties of time--ordered products}{se2}

Let us briefly recall some basic facts from
the
causal perturbation theory (\cite{EG}).
In this approach one constructs
time--ordered products of fields
$$
T_n \bigl(\Theta_1 (\x_1) \cdots \Theta_n (\x_n)\bigr),
$$
where $\Theta_j(\x)$ are local free field polynomials like
$\varphi(\x)$, $\NPR{\varphi^2} (\x)$, $\NPR{\di^{\mu} \varphi \, \di_{\mu} \varphi} (\x)$, $\dots$
(i.e., $\Theta_j (\x)$ are composite fields of free fields).
The main axioms for these time--ordered products are the following:

\begin{itemize}
\item[$(T_0)$]
\textit{Domain}: $T_n \bigl(\Theta_1 (\x_1) \cdots \Theta_n (\x_n)\bigr)$ are operator valued distributions
acting on \textit{invariant} domain that
contains the domain of the Wightman fields;
\item[$(T_1)$]
\textit{Permutation symmetry}:
$$
T_n \bigl(\Theta_1 (\x_1) \cdots \Theta_n (\x_n)\bigr)
\, = \,
(-1)^{\varepsilon} \,
T_n \bigl(\Theta_{\sigma_1} (\x_{\sigma_1}) \cdots \Theta_{\sigma_n} (\x_{\sigma_n})\bigr)
$$
for every permutation $(\sigma_1,\dots,\sigma_n)$ of $(1,\dots,n)$,
where $\varepsilon$ $(=$
$\varepsilon \bigl(\sigma;$ $\Theta_1,$ $\dots,$ $\Theta_n\bigr))$
is the \textit{fermionic} parity of the permutation $\sigma$
for the given set of local Wick polynomials $\Theta_1,$ $\dots,$ $\Theta_n$.
\item[$(T_2)$]
\textit{Covariance}:
\beqa
\podr
U_g T_n \bigl(\Theta_1 (\x_1) \cdots \Theta_n (\x_n)\bigr) U_g^{-1}
\nn \podr = \,
T_n \Bigl(\bigl((\pi(g)^{-1}\Theta_1\bigr) (g\x_1) \cdots \bigl(\pi(g)^{-1}\Theta_n\bigr) (g\x_n)\Bigr)
\nonumber
\eeqa
for every Poincar\'e transformation $g$.\footnote{%
The translations act trivially on the fields, i.e.,
$(\pi(g)\Theta)(\x)$ $=$ $\Theta(\x)$.}
\item[$(T_3)$]
\textit{Causality}:
\beqa
\podr
T_{m+n} \bigl(\Theta_1 (\x_1) \cdots \Theta_n (\x_n)\bigr)
\nn \podr = \
T_m \bigl(\Theta_1 (\x_1) \cdots \Theta_m (\x_m)\bigr)
\,
T_n \bigl(\Theta_{m+1} (\x_{m+1}) \cdots \Theta_{m+n} (\x_{m+n})\bigr)
\nonumber
\eeqa
in the domain\footnote{%
The relation $\x \AFTR \y$ means that $\x \neq \y$ and $\x$ is not contained in the past of $\y$,
i.e., $\x$ $\notin$ $\y-\overline{V}_+$,
where $\overline{V}_+$ is the closure of the open future light--cone $V_+$.}
$\x_j \AFTR \x_{m+k}$ for $j=1,\dots,m$, $k=1,\dots,n$.
In particular, $T_1 \bigl((\Theta (\x)\bigr)$ $=$ $\Theta (\x)$.
\item[$(T_4)$]
\textit{Causal Wick expansion}:
\beqa\label{WTF}
\hspace{-25pt}
T_n \bigl(\Theta_1 (\x_1) \cdots \Theta_n (\x_n)\bigr)
\, = \podr
\mathop{\sum}\limits_{\Rr_1,\dots,\Rr_n}
\frac{1}{\Rr_1! \cdots \Rr_n!} \,
\lvac \hspace{1pt}
T_n \bigl(\Theta_1^{(\Rr_1)} (\x_1) \cdots \Theta_n^{(\Rr_n)} (\x_n)\bigr)
\rvac
\nn \podr \times \,
\NPR{\Phi^{\Rr_1} (\x_1) \cdots \Phi^{\Rr_n} (\x_n)} \,,
\eeqa
where we are using the same ``superquadri--index'' notations like in the Epstein--Glaser paper
\cite{EG}
(their notation $\NPR{\mathrm{A} (\x)^{\rr}}\,$)\footnote{%
The right hand side is well defined due to the zeroth theorem of Epstein--Glaser (\cite{EG}).}.
\item[$(T_5)$]
\textit{Scaling degree}:
there is a grading on the space of composite fields provided by the
so called scaling dimension $\text{dim }\Theta$ and
\beqa
&
\text{The Steinmann scaling degree of }
\lvac \hspace{1pt}
T_n \bigl(\Theta_1 (\x_1) \cdots \Theta_n (\x_n)\bigr)
\rvac
& \nn & \leqslant \,
\text{dim }\Theta_1  + \cdots + \text{dim }\Theta_n \,.
& \nonumber
\eeqa
\item[$(T_6)$]
\textit{Unitarity}:
\beqa
\podr
\hspace{-25pt}
T_n \bigl(\Theta_1 (\x_1)^{+} \cdots \Theta_n (\x_n)^{+}\bigr)^{+}
\, = \,
\mathop{\sum}\limits_{m \, \geqslant \, 0} \,
\mathop{\sum}
\limits_{\text{\scriptsize
\(\begin{array}{c}
\text{ordered partitions}
\\
(S_1,\dots,S_m)
\text{ of}
\\
\{1,\dots,n\}
\end{array}\)}}
(-1)^{n+m} \ \,
T_{S_1} \cdots  T_{S_m} \,,
\nonumber
\eeqa
where
$T_{S_j}$ $:=$ $T_{|S_j|} \Bigl(\mathop{\prod}\limits_{k \, \in \, S_j} \Theta_j (\x_j)\Bigr)$
and $(\cdot)^+$ stands for the Hermitian conjugation.
\end{itemize}

There are some further requirements to the time--ordered products,
like the action Ward identity.
We shall not consider the latter
at this stage but shall make some comments on it in
Sect.~\ref{se6.2}.
Let us also point out that we would like to have the time--ordered products
$T_n \bigl(\Theta_1 (\x_1)$ $\cdots$ $\Theta_n (\x_n)\bigr)$
constructed for every set of composite fields
$\Theta_1 (\x),$ $\dots,$ $\Theta_n (\x)$,
and this also includes the multilinearity condition
mentioned
in \cite[Sect.~4.1]{S2}.

\Section{Sketch of the construction of time--ordered products
by re\-n\-o\-rmalization maps}{se2a}

\noindent
In this section we shall draw in sketch our ideas
for constructing time--ordered products.
We shall explain them more precisely in the subsequent sections.

So, our idea for a construction of time--ordered products,
which is alternative to the old Epstein--Glaser procedure,
is to use
directly
the
the causal Wick expansion
(\ref{WTF}).
In other words, we would like to set
\beqa\label{WTFM}
\hspace{-0pt}
T_n \bigl(\Theta_1 (\x_1) \cdots \Theta_n (\x_n)\bigr)
= \! \podr
\mathop{\sum}\limits_{\Rr_1,\dots,\Rr_n}
\frac{1}{\Rr_1! \cdots \Rr_n!} \,
R_n \Bigl(
\lvac
\Theta_1^{(\Rr_1)} (\x_1) \cdots \Theta_n^{(\Rr_n)} (\x_n)
\rvac
\Bigr)
\nn \podr \times \,
\NPR{\Phi^{\Rr_1} (\x_1) \cdots \Phi^{\Rr_n} (\x_n)} \ ,
\eeqa
where $R_n$ are suitable linear maps
$$
\left\{\hspace{-2pt}
\begin{array}{c}
\text{algebra {$\PA_n$} of non globally defined} \\
\text{$n$--point {\it complex analytic} functions}
\end{array}
\hspace{-2pt}\right\}
\hspace{5pt}
\raisebox{-3pt}{$\mathop{\text{\Huge $\to$}}\limits^{\raisebox{3pt}{\large $\RMA_n$}}$}
\hspace{5pt}
\left\{\hspace{-2pt}
\begin{array}{c}
\text{space of globally} \\
\text{defined distributions}
\end{array}
\hspace{-2pt}\right\} .
$$
The problem then is what are the properties of the maps $R_n$, which will ensure
the properties $(T_0)$--$(T_6)$.
In \cite{N1} we have introduced such maps
for Euclidean Green functions
and called them renormalization maps.
In Sect.~\ref{se5} we shall give a modification of this theory on Minkowski space
and shall show that they provide the main properties of
the
time--ordered products.

At this point we shall only describe
the algebras of regular functions on which we apply the renormalization maps.
Thus, this is a sequence of algebras $\PA_2$, $\PA_3$, $\dots$, $\PA_n$, $\dots$
etc., where $\PA_n$ is an algebra of translation
invariant complex analytic functions of $n$ vector arguments
belonging to the \textit{symmetrized extended $n$--point backward tube}.\footnote{%
The
symmetrized extended $n$--point backward tube
is the domain of analyticity of the $n$--point Wightman functions, which consists
of the orbit of the $n$--point backward tube
$$
\Bigl\{(\x_1,\dots,\x_n) \, \in \, M^{\times n} + i M^{\times n} \, : \,
\x_k-\x_{k+1} \, \in \, M - i V_+ \text{ for } k = 1,\dots,n \Bigr\}
$$
($V_+$ being the open future light--cone)
under the action
of the complexified Lorentz group and the permutation group \cite[IV.5]{J}.}
These functions one can think of as coming from
\textit{analytically extended} Wightman functions
of composite fields.
More precisely, we assume that the
algebra $\PA_n$ is linearly spanned by all finite
linear combinations of products of the form
$$
G \, = \,
\mathop{\prod}
\limits_{1 \, \leqslant \, j \, < \, k \, \leqslant \, n} \,
G_{jk} \bigl(\x_j-\x_k\bigr)
, \quad
$$
where
$\x_k$ $=$ $(x_k^0,\dots,x_k^{D-1})$ are vectors belonging, in general, to the complexified
Min\-ko\-w\-s\-ki space $M+iM$ $\equiv$ $\R^{D-1,1}$ $+$ $i\R^{D-1,1}$
and
the functions $G_{jk}(\x)$ belong to the
algebra $\PA_2$.
The latter algebra, $\PA_2$, is supposed to be a subalgebra of
the algebra of analytic functions
on the extended backward tube in $M+iM$.
One can think of the algebra $\PA_2$ as an
algebra containing the propagators of the theory.
In this way the algebras $\PA_n$ for $n>2$ are
entirely determined by the algebra $\PA_2$.
To retain the permutation symmetry on the algebras $\PA_n$
it is also convenient to continue to consider them
as algebras of regular functions on Euclidean configuration spaces like in \cite{N1}
but now we assume in addition that they also posses certain analytic continuations
mentioned above.

We retain
the technical assumption for the algebra $\PA_2$
(and thus for all other $\PA_n$) that it is closed
with respect to multiplication of its elements by
polynomials as well as with respect to applying derivatives.

We
conclude
this section
with a more suitable form of the
causal Wick expansion
\beqa\label{WTFN}
\podr
T_n
\bigl(
\Theta_1 (\x_1) \cdots \Theta_n(\x_n)
\bigr)
\nn
\podr = \,
R_n \Biggl(
\mathop{\prod}\limits_{1 \, \leqslant \, j \, < \, k \, \leqslant \, n}
\exp \Biggl(
\mathop{\sum}\limits_{\rR,\sS} C_{\rR,\sS} \bigl(\x_j-\x_k\bigr)
\frac{\di}{\di \varphi_{\rR} (\x_j)} \frac{\di}{\di \varphi_{\sS} (\x_k)}
\Biggr)
\Biggr)
\nn \podr \hspace{13pt} \times \,
:\!\Theta_1 (\x_1) \cdots \Theta_n(\x_n)\!: \,,
\eeqa
where\
\(
\varphi_{\rR} (\x) :=
\di_{\x}^{\rR} \varphi (\x)
, \
C_{\rR,\sS} \bigl(\x_1-\x_2\bigr) :=
\di_{\x_1}^{\rR} \di_{\x_2}^{\sS}
\lvac \varphi(\x_1) \varphi(\x_2) \rvac
\)\
and $\di_{\x}^{\rR}$ $=$
$\Bigl(\frac{\di}{\di x^0}\Bigr)^{r_0}$ $\cdots$ $\Bigl(\frac{\di}{\di x^{D-1}}\Bigr)^{r_{D-1}}$.
The formula (\ref{WTFN}) is understood in the following way.
The product
\beq\label{DOP}
\mathop{\prod}\limits_{1 \, \leqslant \, j \, < \, k \, \leqslant \, n}
\exp \Biggl(
\mathop{\sum}\limits_{\rR,\sS} C_{\rR,\sS} \bigl(\x_j-\x_k\bigr)
\frac{\di}{\di \varphi_{\rR} (\x_j)} \frac{\di}{\di \varphi_{\rR} (\x_k)}
\Biggr)
\eeq
in the argument of $R_n$ is considered, after expanding the exponents,
as a differential operator in $\frac{\di}{\di \varphi_{\rR} (\x_j)}$,
which acts on
$:\!\!\Theta_1 (\x_1)$ $\cdots$ $\Theta_n(\x_n)\!\!:$,
where every $\Theta_j (\x_j)$ is considered as a polynomial in
$\{\varphi_{\rR} (\x_j)\}_{\rR}$
for $j=1,\dots,n$.
The coefficients of so obtained differential operator (\ref{DOP})
are elements of the algebra $\PA_n$ on which we then apply the renormalization map $R_n$.

Let us mention that the formula (\ref{WTFN}) is known in the literature also in the form
that uses variational derivatives:
\beqa
\podr
\hspace{-20pt}
T_n
\bigl(
\Theta_1 (\x_1) \cdots \Theta_n(\x_n)
\bigr)
\nn
\podr
\hspace{-20pt}
= \,
R_n \Biggl( \hspace{-2pt}
\exp \Biggl( \hspace{-1pt} \frac{1}{2}
\int \mathrm{d}\z \, \mathrm{d}\w \,
C \bigl(\z-\w\bigr) \,
\frac{\delta}{\delta \varphi (\z)} \frac{\delta}{\delta \varphi (\w)}
\hspace{-1pt} \Biggr)
\hspace{-2pt} \Biggr)
:\!\Theta_1 (\x_1) \cdots \Theta_n(\x_n)\!: \,,
\nonumber
\eeqa
where $C(\x-\y)= \lvac \varphi (\x) \varphi (\y) \rvac$.

\Section{More precise formulation of time--ordered products}{se3}

Let $F$ be the linear span of the generating fields $\varphi_A$ in our theory.
In other words, $F$ is a vector space equipped with a basis $\varphi_A$
that is in one--to--one correspondence with the basic Wightman fields\footnote{%
With a slight abuse of the notations we shall use the one and the same letters
$\varphi_A$ in the two cases.}
$\varphi_A (\x)$.
In particular, $F$ is some representation of the Lorentz group,
which in general is a sum of irreducible representations.
The space $F$ is $\Z_2$--graded (bosons and fermions).
We work with \textit{complex} fields.
Thus, in what follows $F$ and all the vector spaces and algebras
will be considered over the field of complex numbers $\C$.

The space of the derivative fields is
$$
\FSPO \, = \, \C [\di] \otimes F \,,
$$
which we shall also write as
$\FSPO = \C [\di] F$.
Here, $\di$ $:=$ $(\di_0,$ $\dots,$ $\di_{D-1})$ $\equiv$
$\Bigl(\frac{\di}{\di x^0},$ $\cdots,$ $\frac{\di}{\di \x^{D-1}}\Bigr)$
and so, $\FSPO$ is just the linear span of all derivative fields of the form
$$
\bigl(\di^{\rR} \varphi_A\bigr) (\x) \, \equiv \, \varphi_{A,\rR} (\x)
\, := \,
\bigl(\di_0^{r_0} \cdots \di_{D-1}^{r_{D-1}} \varphi_A\bigr) (\x),
$$
where $\rR$ $=$ $(r_0,\dots,r_{D-1})$ is a multiindex.
The space of all composite fields is then introduced as
the free graded--commutative algebra over $\FSPO$, i.e.,
$$
\FSP_{\INFt}
\, = \,
\C \bigl[\FSPO\bigr] \,.
$$
In this way, $\FSP_{\INFt}$ is also a differential algebra
with even derivatives $\di_0,$ $\dots,$ $\di_{D-1}$ (with respect to the $\Z_2$--grading).

Thus, we assumed that the algebra $\FSP_{\INFt}$ consists of
``\textit{off--shell}'' fields $\Theta$,
i.e. it is not factorized by any ideal at the first step.
Later, the elements $\Theta$ $\in$ $\FSP_{\INFt}$ will
be mapped to the space of composite fields $\Theta (\x)$ of free fields
and their derivatives, which act on a (dense domain in a) Hilbert space.
Let us also point out that the product in the algebra $\FSP_{\INFt}$
does not correspond to the product in the operator algebra
of quantum fields, but it rather corresponds to the pointwise normal product.\footnote{%
Our notation $\FSP_{\INFt}$ should correspond to $\mathcal{P}^{\oplus}$
from \cite{S2}.}

For the axiom $(T_5)$ we also need a grading ``$\text{dim}$'' on $\FSP_{\INFt}$
in such a way that it becomes a graded commutative differential
algebra with derivatives $\{\di_{\mu}\}_{\mu \, = \, 0}^{D-1}$ of degree~$1$.
The latter implies that the grading function on $\FSP_{\INFT}$ is completely determined
by its restriction on the space $F$: for instance,
$\text{dim} \, (\varphi_A \di_{\mu} \varphi_B)$ $=$
$1$ $+$ $\text{dim} \, \varphi_A$ $+$ $\text{dim} \, \varphi_B$.
We assume that the $\Z_2$--grading function on $\FSP_{\INFt}$ is expressed
by the grading function $\text{dim}$ as
$2 \, \text{dim}$ $(\text{mod} \, 2)$.
We also require that
\beqa\label{COMP}
&
\text{The Steinmann scaling degree of }
\lvac \hspace{1pt}
\varphi_A (\x_1) \varphi_B (\x_2)\bigr)
\rvac
& \nn &
\leqslant \, \text{dim} \, \varphi_A + \text{dim} \, \varphi_B \,.
&
\eeqa

Next, let us consider for every positive integer $j \in \N \equiv \{1,2,\dots\}$,
a copy of $\FSP_{\INFt}$ denoted by $\FSP_{\INFT \{j\}}$ together with an
identification
\beq\label{IDE}
\FSP_{\INFt} \cong \FSP_{\INFT \{j\}} : \Theta \to (\Theta)_j \,.
\eeq
The elements $(\Theta)_j$ of the space $\FSP_{\INFT \{j\}}$
will further play the role of local fields $\Theta (\x_j)$ in the variable $\x_j$.
Finally, let us introduce the spaces of all polylocal (off--shell) fields
$\FSP_{\INFT S}$:
$$
\FSP_{\INFT S} \, = \, \bigotimes_{j \, \in \, S} \, \FSP_{\INFT \{j\}}
$$
for every finite nonempty subset $S \subset \N$.
Thus, $\FSP_{\INFT S}$ is an algebra,\footnote{%
Again the product in this algebra $\FSP_{\INFT S}$
does not correspond to the product in the operator algebra
of quantum fields.}
which is the tensor product of the \textit{graded} differential algebras
$\{\FSP_{\INFT \{j\}}\}_{j \, \in \, S}$.
The algebras $\FSP_{\INFT S}$ naturally form an \textit{inductive} system
(i.e., if $S' \subseteq S''$ then $\FSP_{\INFT S'}$ $\hookrightarrow$ $\FSP_{\INFT S''}$)
and we introduce also the inductive limit
$$
\FSP_{\INFT \N} \, = \, \mathop{\lim}\limits_{\raisebox{5pt}{$\longrightarrow$}} \, \FSP_{\INFT S}
\, \equiv \, \mathop{\bigcup}\limits_{S \hspace{-4pt} \mathop{\subset}\limits_{\text{finite}} \hspace{-4pt} \N} \, \FSP_{\INFT S}\,.
$$
(Thus, we identify $\FSP_{\INFT \{j\}}$ and $\FSP_{\INFT S}$
with subalgebras in $\FSP_{\INFT \N}$.)

The elements of $\FSP_{\INFT \N}$ will be mapped later either to
sums of
normal products
$:\hspace{-3pt}\Theta_1 (\x_1)$ $\cdots$ $\Theta_n (\x_n)\hspace{-3pt}:$,
or to time--ordered products
$T \bigl(\Theta_1 (\x_1)$ $\cdots$ $\Theta_n (\x_n)\bigr)$.

So, the algebra $\FSP_{\INFT \N}$ is the
free (graded--)commuta\-ti\-ve algebra with generators
$\{\varphi_{A,\rR,j}\}_{A,\,\rR,\,j}$
(corresponding to the fields
$(\di^{\rR}\varphi_A)(\x_j)$).
This algebra has various structures on it
(several products, derivatives, coproducts, structure of a jet algebra, \dots)
and it is very interesting to find the
interplay
between them.
We shall not investigate here the possible algebraic structures on $\FSP_{\INFT \N}$
but we shall mention only one, which will be used further.
This is the permutation symmetry:
for every bijection $\sigma : \N \cong \N$ there is a natural action
$\sigma : \FSP_{\INFT \N} \to \FSP_{\INFT \N}$, which is a representation
of the (infinite) permutation group $\Ss (\N)$ $\ni$ $\sigma$.
To write this action explicitly, note that the algebra $\FSP_{\INFT \N}$
is linearly spanned by products of the form
\beq\label{MF1}
\mathop{\prod}\limits_{j \, \in \, \N} \, \Theta_j
\eeq
where $\Theta_j \in \FSP_{\INFT \{j\}}$ (the $j$-th copy of $\FSP_{\INFt}$)
and only finitely many $\Theta_j$ are different from 1.
Then the action $\sigma : \FSP_{\INFT \N} \to \FSP_{\INFT \N}$ is defined by the formula
$$
\sigma \Biggl(\mathop{\prod}\limits_{j \, \in \, \N} \, \Theta_j\Biggr)
\, = \,
\mathop{\prod}\limits_{j \, \in \, \N} \,
\Theta_{\sigma(j)}
\,.
$$
(In particular, in the purely bosonic case the above action of $\Ss(\N)$ is trivial.)

In order to introduce the linear maps from $\FSP_{\INFT \N}$
into the space of operator valued distributions,
which give the normal and time--ordered products
let us denote for every finite $S \subset \N$:
$$
\DP_S \, := \, \DP \bigl(M_S\bigr) \,,
$$
where
$$
M_S \, := \, M^S \bigl/ M
$$
is the space of $S$--point configurations $\bigl(\x_j\bigr)_{j \, \in \, S}$ of vectors in $M$
modulo translations
(the elements of $M_S$ will be denoted by
$$
\bigl[\x_j\bigr]_{j \, \in \, S} \, := \,
\bigl(\x_j\bigr)_{j \, \in \, S} \text{ mod } M \,).
$$
Thus, $\DP_S$ is
the space of all translation invariant ``$S$--point'' distributions
over the Minkow\-s\-ki space $M$ (i.e., distributions whose vector arguments $\x_j$
are indexed by elements $j \in S$).
Also, let us denote
$$
\HDP_S \, := \, \DP \bigl(M^S, Op (\DOM)\bigr)
$$
the space of all $S$--point operator valued distributions acting on an invariant dense domain
$\DOM$ in a Hilbert space.
Let us introduce again the inductive limits
$$
\DP_{\N} \, = \,
\mathop{\bigcup}\limits_{S \hspace{-4pt} \mathop{\subset}\limits_{\text{finite}} \hspace{-4pt} \N} \, \DP_S
,\quad
\HDP_{\N} \, = \,
\mathop{\bigcup}\limits_{S \hspace{-4pt} \mathop{\subset}\limits_{\text{finite}} \hspace{-4pt} \N} \, \HDP_S \,,
$$
and the actions of the permutation group $\Ss (\N)$ $(\ni \sigma)$
on the spaces $\DP_{\N}$ and $\HDP_{\N}$, which are given by the formula
$$
(\sigma f) (\x_1,\x_2,\dots) \, = \, f\bigl(\x_{\sigma^{-1}(1)},\x_{\sigma^{-1}(2)},\dots\bigr) \,.
$$

Now, in a Wightman theory of a system of free fields $\{\varphi_A\}$ we have linear maps
$$
N : \FSP_{\INFT \N} \to \HDP_{\N}
\, , \quad
T : \FSP_{\INFT \N} \to \HDP_{\N}
\, .
$$
The first of them is the normal product:
$$
N \Biggl(\mathop{\prod}\limits_{j \, \in \, \N} \, \Theta_j\Biggr)
\, := \
\NPR{\Theta_1 (\x_1) \, \Theta_2 (\x_2) \cdots}
$$
and it has a canonical realization so that
we shall not characterize it here axiomatically.
The second map $T$ gives the time ordered products:
$$
T \Biggl(\mathop{\prod}\limits_{j \, \in \, \N} \, \Theta_j\Biggr)
\, := \
T \bigl(\Theta_1 (\x_1) \, \Theta_2 (\x_2) \cdots\bigr)
$$
and it does not have a canonical construction
but is defined up to
a
renormalization ambiguity.

It is easy to
reformulate
the axiomatic properties $(T_0)$--$(T_6)$ for
the time ordered products in terms of the maps $N$ and $T$.
In particular,
both maps, $N$ and $T$, have the permutation symmetry:
$$
\sigma \circ N \, = \, N \circ \sigma
\, , \quad
\sigma \circ T \, = \, T \circ \sigma
$$
for $\sigma \in \Ss (\N)$.

The causality reads
\beq\label{CAUS}
T\bigl(\Theta_{S'} \,\Theta_{S''}\bigr)\vrestr{12pt}{\CSL_{S;S'}}
\, = \,
T\bigl(\Theta_{S'}\bigr) \, T\bigl(\Theta_{S''}\bigr)\vrestr{12pt}{\CSL_{S;S'}}
\eeq
where $S = S' \, \dot{\cup} \, S''$ (disjoint union),
$\Theta_{S'}$ $\in$ $\FSP_{S'}$ and $\Theta_{S''}$ $\in$ $\FSP_{S''}$,
and $\CSL_{S;S'}$ is the open region:
$$
\CSL_{S;S'} \, := \,
\Bigl\{
\bigl[\x_j\bigr]_{j \, \in \, S} \, : \,
\x_{j'} \, \AFTR \, \x_{j''} \text{\, for\, }
j' \, \in \, S' \text{\, and\, } j'' \, \in \, S''
\Bigr\} \,
$$
(the relation $\x_{j'} \AFTR \x_{j''}$ stands for $\x_{j''}$ $\notin$ $\x_{j'}-\overline{V}_+$).

The unitarity reads
\beqa\label{UNI}
\podr
T \bigl(\Theta_S^{\Cnj}\bigr)^{+}
\, = \,
\mathop{\sum}\limits_{m \, \geqslant \, 0} \,
\mathop{\sum}
\limits_{\text{\scriptsize
\(\begin{array}{c}
\text{ordered partitions}
\\
(S_1,\dots,S_m)
\text{ of } S
\end{array}\)}}
(-1)^{|S|+m} \ \,
T \bigl(\Theta_{S_1}\bigr) \cdots  T \bigl(\Theta_{S_m}\bigr) \qquad
\eeqa
where $\Theta_{S_k}$ $\in$ $\FSP_{\INFT S_k}$,
$\Theta_S$ $=$ $\Theta_{S_1}$ $\cdots$ $\Theta_{S_m}$
and $\Theta$ $\mapsto$ $\Theta^{\Cnj}$
is the involutive anti--automorphism of the algebra $\FSP_{\INFT \N}$
generated by the antilinear involution on the space $F$ that represents the field conjugation.

\Section{Causal Wick expansion and causality}{se4}

In order to formulate the
causal Wick expansion
in terms of the notations of Sect.~\ref{se3}
let us first introduce
for every finite subset $S$ $\subset$ $\N$
the algebra $\PA_S$, which is the linear span of all
\beq\label{GSF}
G \, = \,
\mathop{\prod}
\limits_{\mathop{}\limits^{j,\, k \, \in \, S}_{j \, < \, k}} \,
G_{jk} \bigl(\x_j-\x_k\bigr)
, \quad
G_{jk} (\x) \in \PA_2.
\eeq
The system of algebras $\{\PA_S\}_S$ is again an inductive system and we set
$$
\PA_{\N} \, = \,
\mathop{\bigcup}\limits_{S \hspace{-4pt} \mathop{\subset}\limits_{\text{finite}} \hspace{-4pt} \N} \, \PA_S
\, = \,
\mathop{\bigcup}\limits_{n \, = \, 1}^{\infty} \, \PA_n \,.
$$
As in the above cases of such inductive limits we have also a natural action
$$
\sigma : \PA_{\N} \to \PA_{\N}
$$
of $\Ss (\N)$ $\ni$ $\sigma$.

The system of renormalization maps $\RMA_S$ $:$ $\PA_S$ $\to$ $\RMA_S$
(to be defined in Sect.~\ref{se5})
will be consistent with the above inductive limits and thus, they will
induce a linear map
\beq\label{RMAN}
\RMA : \PA_{\N} \to \DP_{\N}
\ ,\qquad
\RMA \bigl(\PA_S \bigr) \, \subseteq \, \DP_S
\,.
\eeq

Now, the
causal Wick expansion
(\ref{WTFN})
has also more compact form in the above notations,
\beq\label{WTE1}
T\bigl(\Theta_S\bigr) \, = \,
\RMA \bigl(\GH_S\bigr) N\bigl(\Theta_S\bigr)
\,,
\eeq
where $\Theta_S \in \FSP_S$,
and $\GH_S$ is an operator
\beq\label{gh1}
\GH_S : \PA_S \otimes \FSP_S \to \PA_S \otimes \FSP_S
\eeq
defined by
\beq\label{gh2}
\GH_S \, = \, \mathop{\prod}\limits_{\mathop{}\limits^{j,\, k \, \in \, S}_{j \, < \, k}}
\GH_{jk}
,\quad
\GH_{jk} \, = \,
\exp \Biggl(
\mathop{\sum}\limits_{A,B,\rR,\sS} C_{A,\rR;B,\sS} \bigl(\x_j-\x_k\bigr)
 \otimes
\frac{\di}{\di \varphi_{A,\rR,j}} \frac{\di}{\di \varphi_{B,\sS,k}}
\Biggr)
\eeq
and
\(
C_{A,B,\rR,\sS} \bigl(\x_1-\x_2\bigr) :=
\di_{\x_1}^{\rR} \di_{\x_2}^{\sS}
\lvac \varphi_A(\x_1) \varphi_B(\x_2) \rvac
\in \PA_2
\).
Let us note that the derivations $\frac{\di}{\di \varphi_{A,\rR,j}}$
entering in Eq.~(\ref{gh2}) are the canonical graded derivations
on the algebra $\FSP_{\INFT \N}$ regarded as a free graded commutative algebra
with generators $\{\varphi_{A,\rR,j}\}_{A,\rR,j}$ (these are the ``vertical derivations'').
We also note that equation (\ref{WTE1}) is more correct to write as
\beq\label{WTENN1}
T \bigl(\Theta_S\bigr) \, = \, \text{mult} \circ \bigl(R \otimes N\bigr) \bigl(\GH_S \Theta_S\bigr) \,,
\eeq
where $\text{mult}$ $:$ $\DP_{\N} \otimes \HDP_{\N \, (\text{norm})}$ $\to$ $\HDP_{\N}$
is the operation of multiplication between the distributions in the space $\DP_{\N}$
and the operator--valued distributions in the subspace $\HDP_{\N \, (\text{norm})}$ $\subset$ $\HDP_{\N}$
spanned by the normal products of local fields of different arguments
(this multiplication exists due to the zeroth theorem of Epstein and Glaser \cite{EG}).

\medskp

\begin{Remark}\label{RR3}
The maps $\GH_S$ are consistent with the inductive limit and generate a linear map
$\GH$ $:$ $\PA_{\N} \otimes \FSP_{\INFT \N}$ $\to$ $\PA_{\N} \otimes \FSP_{\INFT \N}$.
Then Eq.~(\ref{WTENN1}) reads
$$
T \, = \, \text{mult} \circ \bigl(R \otimes N\bigr) \circ \GH \,.
$$
We shall work with $\GH_S$ because of the combinatorics related to the causality.
\end{Remark}

\medskp

We have also a
Wick expansion formula
for ordinary products.
In order to formulate it we need additional notations.
Since the functions belonging to $\PA_S$ are analytic functions
on certain $S$--point tube domains we have another linear maps
$\PA_{S}$ $\to$ $\DP_{S}$, the boundary values of analytic functions.
In order to define them we introduce the notion
of an \textit{ordered} set $\VE{S}$
that is a set $S$ $=$ $\{j_1,\dots,j_n\}$ equipped with a total order
$j_1$ $\LE$ $\cdots$ $\LE$ $j_n$ on it.
We shall also write it as
$$
\VE{S} \, = \, \La j_1,\dots,j_n \Ra \,
$$
and $S$ will be called a body of $\VE{S}$.
If $S \subset \N$ we shall consider the order $\LE$ on $S$
as an \textit{independent} structure
on it, which may not coincide with the order $<$ induced by $\N$.
For every ordered set $\VE{S}$
$=$ $\La j_1,\dots,j_n \Ra$
we have
a standard backward\footnote{%
Since we use in (\ref{GSF}) differences of a type left minus right coordinate
then it follows that the boundary values appear
with respect to backward tubes instead of forward.}
tube domain associated to $\VE{S}$,
$$
\TUBE_{\VE{S}}
\, := \Bigl\{[\x_j]_{j \, \in \, S} \, \in \, M_S + i M_S \, : \,
\x_{j_k}-\x_{j_{k+1}} \, \in \, M - i V_+ \text{ for } k = 1,\dots,n \Bigr\}
$$
($V_+$ being the open forward light--cone in $M$)
and then we define a boundary value map
$$
\BV_{\VE{S}} : \PA_S \to \DP_S
$$
with respect to this tube $\TUBE_{\VE{S}}$.
Thus, the linear maps $\BV_{\VE{S}}$ will produce the Wightman functions in the theory.

Now, the
Wick formula
for ordinary products is
\beq\label{WTEW}
N\bigl(\Theta_{j_1}\bigr) \cdots N\bigl(\Theta_{j_1}\bigr)
\, = \,
\BV_{\VE{S}} \bigl(\GH_S\bigr) \ N \bigl(\Theta_S\bigr)
\eeq
($\VE{S}$ $=$ $\La j_1,\dots,j_n \Ra$).
Again Eq.~(\ref{WTEW})
is understood as
\beq\label{WTEW1}
N\bigl(\Theta_{j_1}\bigr) \cdots N\bigl(\Theta_{j_1}\bigr)
\, = \,
\text{mult} \circ
\bigl(\BV_{\VE{S}} \otimes N\bigr) \bigl(\GH_S \Theta_S\bigr) \,.
\eeq

Equation (\ref{WTEW1}) shows that the right hand side of Eq.~(\ref{WTENN1})
always defines an operator $\FSP_{\INFT \N}$ $\to$ $\HDP_{\N}$
since the right hand sides of Eqs.~(\ref{WTEW1}) and (\ref{WTENN1})
can be written as:
$$
\bigl(\BV_{\VE{S}} \otimes \ID\bigr) \circ
\bigl(\ID \otimes N\bigr) \bigl(\GH_S \Theta_S\bigr)
\quad \text{and} \quad
\bigl(\RMA \otimes \ID\bigr) \circ
\bigl(\ID \otimes N\bigr) \bigl(\GH_S \Theta_S\bigr)
\,,
$$
respectively, and the map $\BV_{\VE{S}}$ is an \textit{injection}.
Hence, our idea to construct time--ordered products by
Eq.~(\ref{WTFN}) or Eq.~(\ref{WTFM})
is correct even if we start with off--shell fields:
the corresponding new formula is (\ref{WTENN1}) and it will
preserve
the kernel in the passage from off--shell fields to on--shell fields.

An important consequence of Eq. (\ref{WTEW}) is
\beq\label{WTE2}
N \bigl(\Theta_{S'}\bigr) N \bigl(\Theta_{S''}\bigr) \, = \,
\BV_{\VE{S}} \bigl(\GH_{S',\,S''}\bigr) \ N \bigl(\Theta_S\bigr)
\,,
\eeq
where for a disjoint union $S$ $=$ $S' \, \dot{\cup} \, S''$
of ordered sets $\VE{S}'$ and $\VE{S}''$ we equip $S$ with an
order $S'$ $\LE$ $S''$ and introduce the splittings
$\Theta_S$ $=$ $\Theta_{S'}\Theta_{S''}$ and
\beqa
\podr \hspace{100pt}
\GH_S \, = \, \GH_{S'} \, \GH_{S''} \, \GH_{S',\,S''} \,,
\nn \podr
\GH_{S'} \, = \, \mathop{\prod}\limits_{\mathop{}\limits^{j,\, k \, \in \, S'}_{j \, \LE \, k}}
\GH_{jk}
\, , \quad
\GH_{S''} \, = \, \mathop{\prod}\limits_{\mathop{}\limits^{j,\, k \, \in \, S''}_{j \, \LE \, k}}
\GH_{jk}
\, , \quad
\GH_{S',\,S''} \, = \, \mathop{\prod}\limits_{\mathop{}\limits^{j \, \in \, S'}_{j \, \in \, S''}}
\GH_{jk}
\,.
\nonumber
\eeqa

The causality relation (\ref{CAUS}) will follow now from Eq. (\ref{WTE2})
if $\RMA$ satisfies the following equation:
$$
\RMA \bigl(\GH_S\bigr) \vrestr{12pt}{\CSL_{S;S'}} \, = \,
\RMA \bigl(\GH_{S'}\bigr) \, \RMA \bigl(\GH_{S''}\bigr) \, \BV_{\VE{S}} \bigl(\GH_{S',\,S''}\bigr)
\vrestr{12pt}{\CSL_{S;S'}}
\,.
$$
This will be implied by the recursive axiomatic property $(r4)$ of the renormalization maps $\RMA$
introduced in the next section.
Before we pass to the definition of renormalization maps we shall show
how Eq.~(\ref{WTE2}) implies (\ref{CAUS}).
{\samepage
On $\CSL_{S;S'}$ we have:
\beqa
T\bigl(\Theta_S\bigr)
\, = \podr
\RMA \bigl(\GH_S\bigr) N\bigl(\Theta_S\bigr)
\, = \,
\RMA \bigl(\GH_{S'}\bigr) \RMA \bigl(\GH_{S''}\bigr) \, \BV_{\VE{S}} \bigl(\GH_{S',\,S''}\bigr)
\, N\bigl(\Theta_S\bigr)
\nn
\, = \podr
\RMA \bigl(\GH_{S'}\bigr) \RMA \bigl(\GH_{S''}\bigr)
N \bigl(\Theta_{S'}\bigr) N \bigl(\Theta_{S''}\bigr)
\, = \,
T\bigl(\Theta_{S'}\bigr) T\bigl(\Theta_{S''}\bigr)
\nonumber
\eeqa
($\Theta_S := \Theta_{S'} \Theta_{S''}$).}

\Section{Theory of renormalization maps on Minkowski space}{se5}

Let us now consider the modification of the theory of renormalization maps
on the Minkowski space.
We introduced above an improved version of them
replacing the whole system $\{\RMA_S\}_S$ with a single linear map
$$
\RMA : \PA_{\N} \to \DP_{\N} \,
$$
(and so, $\RMA_S$ $\equiv$ $\RMA\vrestr{10pt}{\PA_S}$).
In order to make sure that this be possible
we need to assume that the system
of renormalization maps $\RMA_S : \PA_S \to \DP_S$ is
consistent with the inductive limits.
This is then equivalent to the enhanced version of the requirement $(r1)$,
which we discussed in Remarks 2.1--2.3 of \cite{N1}.

Thus, the modified axiomatic requirements on the map $\RMA$ are the following.
The conditions $(r1)$--$(r3)$ remains essentially the same
on the Minkowski and the Euclidean space:

\medskip

$(r1)$ \textit{Permutation symmetry.}
For every $\sigma \in \Ss (\N)$ we require
$\sigma \circ \RMA$ $=$ $\RMA \circ \sigma$.

\medskip

$(r2)$ \textit{Preservation of the filtrations.}
The scaling degree does not increase
$\scdeg \, R (G)$ $\leqslant$ $\scdeg \, G$.

\medskip

$(r3)$ \textit{Commutativity with multiplication by polynomials.}
If $p$ is a polynomial on $M_S$ ($S \subset \N$) then
$R (p \hspace{1pt} G)$ $=$ $p \, R (G)$.

\medskip

\begin{Remark}\label{Rm}
Property $(r3)$ might look artificial from physical point of view
and in fact it is not necessary for the construction of the time--ordered products.
In paper \cite{N1} this property  plays a crucial role for the reduction of
the cohomological analysis of the renormalization group
to de Rham cohomologies of configuration spaces.
We considered in $(r3)$ only polynomials
since we wish to work algebraically.
But if we work on manifolds
then it is natural to require commutativity
between the renormalization maps and multiplication
by everywhere smooth functions,
i.e., $R (p \hspace{1pt} G)$ $=$ $p \, R (G)$ for
$p \in \CI \bigl(M_S\bigr)$.
Then the latter property becomes very natural
from geometric point of view since it allows us to make localization
(i.e., to use localization techniques like partition of unity).
\end{Remark}

\medskp

The last condition $(r4)$ needs more essential modification

\medskip

$(r4)$
\textit{Causality.}
For every disjoint union $S$ $=$ $S' \, \dot{\cup} \, S''$ we have
\beq\label{RCAUS}
\RMA \bigl(G_S\bigr) \vrestr{12pt}{\CSL_{S;S'}} \, = \,
\RMA \bigl(G_{S'}\bigr) \, \RMA \bigl(G_{S''}\bigr) \, \BV_{\VE{S}} \bigl(G_{S',\,S''}\bigr)
\vrestr{12pt}{\CSL_{S;S'}}
\eeq
for every $G_S \in \PA_S$ of the form (\ref{GSF}),
where we consider some order on the sets
$S'$ and $S''$ and equip $S$ with the order induced by
$S'$ $\LE$ $S''$; finally, as above, we introduce the splitting
\beqa\label{GSPL}
\podr \hspace{100pt}
G_S \, = \, G_{S'} \, G_{S''} \, G_{S',\,S''} \,,
\nn \podr
G_{S'} \, = \, \mathop{\prod}\limits_{\mathop{}\limits^{j,\, k \, \in \, S'}_{j \, \LE \, k}}
G_{jk}
\, , \quad
G_{S''} \, = \, \mathop{\prod}\limits_{\mathop{}\limits^{j,\, k \, \in \, S''}_{j \, \LE \, k}}
G_{jk}
\, , \quad
G_{S',\,S''} \, = \, \mathop{\prod}\limits_{\mathop{}\limits^{j \, \in \, S'}_{j \, \in \, S''}}
G_{jk}
\,. \qquad
\eeqa
The right hand side of Eq.~(\ref{RCAUS}) is correct due to the following lemma.

\medskip

\begin{Lemma}\label{L1}
The product
$\RMA \bigl(G_{S'}\bigr) \RMA \bigl(G_{S''}\bigr) \BV_{\VE{S}} \bigl(G_{S',\,S''}\bigr)$
exists on
$M_S$.
\end{Lemma}

\medskip

\noindent
\textit{Proof.}
We have to show that the sums of the wave front sets
over\footnote{%
We shall specify in the wave front also the base on which the distribution is
considered since we are using inductive systems of spaces $\DP_S$
$( \, := \DP (M_S),$ $M_S := M^S / M)$
and
a distribution from $\DP_S$ can be also considered in $\DP_{S'}$ for $S \subseteq S'$.}
$M_S$
\beqa\label{WF0}
&
\WF{M_S}{\RMA \bigl(G_{S'}\bigr)}+\WF{M_S}{\RMA \bigl(G_{S''}\bigr)},
& \nn &
\WF{M_S}{\RMA \bigl(G_{S'}\bigr)}+\WF{M_S}{\BV_{\VE{S}} \bigl(G_{S',\,S''}\bigr)},
& \nn &
\WF{M_S}{\RMA \bigl(G_{S''}\bigr)}+\WF{M_S}{\BV_{\VE{S}} \bigl(G_{S',\,S''}\bigr)},
& \nn &
\WF{M_S}{\RMA \bigl(G_{S'}\bigr)}+\WF{M_S}{\RMA \bigl(G_{S''}\bigr)}
+\WF{M_S}{\BV_{\VE{S}} \bigl(G_{S',\,S''}\bigr)}
& \quad
\eeqa
do not intersect the zero section of the cotangent bundle $T^* \bigl(M_S\bigr)$.
This is clear for the first three sums and for the last we first use
a general statement about the wave front of boundary values of analytic functions,
which implies
$$
\BV_{(-)} \bigl(G_{jk}\bigr) \, \subseteq \, M \oplus V_-
$$
where $\BV_{(-)} \bigl(G_{jk}\bigr)$ is the boundary value of $G_{jk} (\x)$
in the backward tube $M+iV_-$, $V_-$ $=$ $-V_+$ (the opened backward light--cone).
It then follows that
\beq\label{WF1}
\WF{M_S}{\BV_{\VE{S}} \bigl(G_{S',\,S''}\bigr)}
\, \subseteq \,
\mathop{\sum}\limits_{\mathop{}\limits^{j \, \in \, S'}_{k \, \in \, S''}}
\bigl(\pi^S_{\{j,k\}}\bigr)^*
\Bigl(M \oplus V_-\Bigr)\,,
\eeq
where $\bigl(\pi^S_{S'}\bigr)^*$ $:$ $T^* \bigl(M_{S'}\bigr)$ $\to$ $T^* \bigl(M_S\bigr)$
is defined for every $S' \subseteq S$ as the pull-back of
the natural projection $\pi^S_{S'}$ $:$ $M_S$ $\to$ $M_{S'}$ $:$
$[\x_j]_{j \, \in \, S}$ $\mapsto$ $[\x_j]_{j \, \in \, S'}$.
Let us combine Eq.~(\ref{WF1}) with
\beqa\label{WF2}
\WF{M_S}{\RMA \bigl(G_{S'}\bigr)}
\, \subseteq \podr
\bigl(\pi^S_{S'}\bigr)^*
\Bigl(\WF{M_{S'}}{\RMA \bigl(G_{S'}\bigr)}\Bigr)
\,, \quad
\nn
\WF{M_S}{\RMA \bigl(G_{S''}\bigr)}
\, \subseteq \podr
\bigl(\pi^S_{S''}\bigr)^*
\Bigl(\WF{M_{S''}}{\RMA \bigl(G_{S''}\bigr)}\Bigr)
\eeqa
and use the splitting
\beq\label{splt}
M_S \cong M_{S'} \hspace{1pt} \oplus \, M_{S''} \hspace{1pt} \oplus \, M : [\x_j]_{j \, \in \, S}
\hspace{1pt} \mapsto \hspace{1pt}
[\x_j]_{j \, \in \, S'} \hspace{1pt} \oplus \, [\x_j]_{j \, \in \, S''} \hspace{1pt} \oplus \,
\bigl(\x_{\max \, S'} - \x_{\min \, S''}\bigr)
.
\eeq
Then
the fibers in the tangent bundle $T \bigl(M_S\bigr)$ spits according to (\ref{splt}),
which then implies a decomposition of the cotangent bundle.
Let $pr_3$ $:$ $T^* \bigl(M_S\bigr)$ $\to$ $T^* \bigl(M_S\bigr)$
be the projection that corresponds to the third summand in Eq.~(\ref{splt})
(i.e., the projection onto
the annihilator of the tangent spaces of the first two summands in Eq.~(\ref{splt})).
it follows that $pr_3$ maps the right hand sides in Eq.~(\ref{WF2}) to the zero section,
and on the other hand, $pr_3$ maps the right hand side of Eq.~(\ref{WF1}) to $M$ $\oplus$ $V_-$.
So, the fourth sum in Eq.~(\ref{WF0}) also does not intersect the zero section
of $T^* \bigl(M_S\bigr)$.$\quad\Box$

\bigskip

This completes the list of modified axiomatic conditions for the renormalization map
$\RMA$
on the Minkowski space.
Let us add one more natural assumption

\medskip

$(r5)$ \textit{Lorentz invariance.}
The map $\RMA$ $:$ $\PA_{\N} \to \DP_{\N}$ intertwines the natural actions
of the Lorentz group on $\PA_{\N}$ and $\DP_{\N}$.

\medskip

As in the Euclidean case, the above invariance is ensured later by the construction
of $\RMA$.

Now, the construction of $\RMA$ follows the same scheme like in the Euclidean case.
First, since the the sets $\CSL_{S;S'}$ form an open covering of $M_S \mz$
we have again inductively defined secondary renormalization maps
$$
\SRMA_S : \PA_S \to \DP_{temp} \bigl(M_S \mz\bigr) \,.
$$
Hence, what remains to do is to compose $\SRMA_S$
$$
\SRMA_S \circ \PRMA_S \, =: \, \RMA_S
$$
with a primary renormalization map\footnote{%
The systems of linear $\PRMA_S$ and $\SRMA_S$ do not form inductive systems.
This is because the system of the intermediate spaces $\DP_{temp} \bigl(M_S \mz\bigr)$
is not an inductive system.}
$$
\PRMA_S : \DP_{temp} \bigl(M_S \mz\bigr) \to \DP_S \,.
$$
Following  \cite{N1}, the linear maps $\PRMA_S$ can be constructed
by means of a larger
system of linear maps $\Prma_N$ $:$ $\DP_{temp} \bigl(\R^N \mz\bigr)$ $\to$ $\DP (\R^N)$,
so that $\PRMA_S$ $\cong$ $\Prma_{D(|S|-1)}$.
The axiomatic properties for $\Prma_N$: $(p1)$--$(p5)$ from \cite{N1}
together with $(p6)$ (\cite[Remark 2.2]{N1}),
remain almost the same
except
the requirement of Euclidean invariance in $(p3)$, which have to be replaced
by the permutation symmetry and Lorentz invariance for $\PRMA_S$.\footnote{%
It is more convenient to exclude from the initial requirements for $\Prma_N$ in
\cite{N1} the property $(p3)$ and just at the end ensure it in the above modified form
for $\PRMA_n$ $\cong$ $\Prma_{D(n-1)}$ by using semi-simplicity of the Lorentz group.}

\medskip

This completes the construction of the renormalization map $\RMA$ on the Minkowski space.

\medskp

\begin{Remark}\label{RM2}
Condition $(r4)$ can be generalized also for arbitrary
\textit{ordered} $S$--partitions.
Theorem 2.9 in \cite{N1}
about changes of renormalization will remain valid in the same form on Minkowski space.
Its proof however should be modified and we have to use now ordered partitions
in order to apply the recursion according to generalized property $(r4)$.
\end{Remark}

\medskp

At the end of this section let us summarize what we have proven
for the construction of time--ordered products.

\medskip

\begin{Theorem}\label{T2}
Let $\RMA : \PA_{\N} \to \DP_{\N}$
be a renormalization map that satisfies
the requirements $(r1)$--$(r5)$ listed above.
Let us define a linear map $T$ $:$ $\FSP_{\INFT \N}$ $\to$ $\HDP_{\N}$ by
Eq.~$($\ref{WTE1}$)$ $($or, Eq. $($\ref{WTENN1}$)$$)$.
Then the map $T$ is well defined and satisfies the properties
$(T_0)$--$(T_5)$ listed in Sect.~\ref{se2}
$($but possibly without the unitarity~$(T_6)$$)$.
\end{Theorem}

\medskip

\noindent
\textit{Final remarks on the proof.}
We have already argued why $(T_0)$--$(T_4)$ are satisfied
and we would like to discuss here condition $(T_5)$,
the scaling degree.
This is ensured by the fact that the operators $\GH_{jk}$ (\ref{gh2}),
and hence, $\GH_S$ (\ref{gh1}),
preserve the filtration on $\PA_{\N} \otimes \FSP_{\INFT \N}$
(this filtration is the induced one from the scaling degree on $\PA_{\N}$
and the grading on $\FSP_{\INFT \N}$: $\leqslant \scdeg \otimes \text{dim}$).
This is simply because the operators
$C_{A,\rR;B,\sS} \bigl(\x_j-\x_k\bigr)$
$\frac{\di}{\di \varphi_{A,\rR,j}}$ $\frac{\di}{\di \varphi_{B,\sS,k}}$
in the arguments of the exponent in
$\GH_{jk}$ have this property:
the derivatives
$\frac{\di}{\di \varphi_{A,\rR,j}}$ $\frac{\di}{\di \varphi_{B,\sS,k}}$
decreases the filtration by
$\text{dim} \, \varphi_{A,\rR,j}$ $+$ $\text{dim} \, \varphi_{B,\sS,k}$,
while $C_{A,\rR;B,\sS} \bigl(\x_j-\x_k\bigr)$
($=$ $\di_{\x_j}^{\rR} \di_{\x_k}^{\sS} \lvac \varphi_A(\x_j)$ $\varphi_B(\x_k) \rvac$)
increases the filtration with the same value.
Note that at this point we also use condition (\ref{COMP})
on the grading function $\text{dim}$,
which implies that
the scaling degree of
$C_{A,\rR;B,\sS} \bigl(\x_j-\x_k\bigr)$
is less than or equal to
$\text{dim} \, \varphi_{A,\rR,j}$ $+$ $\text{dim} \, \varphi_{B,\sS,k}$.

\Section{Unitarity}{se6.1}

As we mentioned in Theorem \ref{T2}
we do not know
whether the properties $(r1)$--$(r5)$ imply the unitarity
condition $(T_6)$ (Eq.~(\ref{UNI}))
for the time--ordered products.
What we can only say now is that a sufficient condition for this would be
the following additional property on the renormalization map $\RMA$:
\beqa\label{er6}
\overline{\RMA \bigl(G_S^{\CNJ}\bigr)}
\, = \,
\mathop{\sum}\limits_{m \, \geqslant \, 0} \,
\mathop{\sum}
\limits_{\text{\scriptsize
\(\begin{array}{c}
\text{ordered partitions}
\\
(S_1,\dots,S_m)
\text{ of } S
\end{array}\)}}
\podr
(-1)^{|S|+m} \ \,
\RMA\bigl(G_{S_1}\bigr) \cdots \bigl(G_{S_m}\bigr)
\nn \podr
\times \
\BV_{\VE{S}_{S_1,\dots,S_m}} \Bigl(G_{\{S_1,\dots,S_m\}}\Bigr)
\eeqa
($G_S \in \PA_S$ has the form (\ref{GSF})).
Here:
the $\CNj$--operation on $G_S^{\CNJ}$ is the ``CPT--operation'',\footnote{%
In particular, this assumes that $\PA_S$ is $\CNj$--invariant,
which thus should be added to the axiomatic requirements on $\PA_S$ (and enough, for $\PA_2$).
The operation $\CNj$ is an anti--automorphism of $\PA_{\N}$.}
$$
G_S^{\CNJ} (\x_{j_1},\dots,\x_{j_n}) \, := \,
\overline{G_S (-\x_{j_1},\dots,-\x_{j_n})}
\qquad (S \, = \, \{j_1,\dots,j_n\})
\,;
$$
the ordered sets $\VE{S}_{S_1,\dots,S_m}$ defining the boundary values in Eq.~(\ref{er6})
is obtained by any orders on $S_k$ and
$S_1$ $\LE$ $\cdots$ $\LE$ $S_m$;
the notation $G_{\{S_1,\dots,S_m\}}$ follows the conventions from \cite[see Eq. (2.14)]{N1},
i.e.,
$$
G_{\{S_1,\dots,S_m\}} \, = \,
\mathop{\prod}\limits_{1 \, \leqslant \, a \, < \, b \, \leqslant \, m} \,
\mathop{\prod}\limits_{\mathop{}\limits^{j \, \in \, S_a}_{k \, \in \, S_b}} \,
G_{jk} (\x_j-\x_k) \,
$$
(which generalizes the notation $G_{S',\,S''}$ (\ref{GSPL})).
Note that the product of the distributions in the right hand side of
Eq.~(\ref{er6}) exists, which can be proven in the same way as Lemma~\ref{L1}.
The unitarity condition (\ref{UNI}) can be derived from Eq. (\ref{er6})
if we use a generalization of Eq.~(\ref{WTE2}) that is
\beq\label{WTE3}
N \bigl(\Theta_{S_1}\bigr) \cdots  N \bigl(\Theta_{S_m}\bigr) \, = \,
\BV_{\VE{S}_{S_1,\dots,S_m}} \Bigl(\GH_{\{S_1,\dots,S_m\}}\Bigr) \,
N \bigl(\Theta_S\bigr)
\eeq
($\Theta_{S_k}$ $\in$ $\FSP_{\INFT S_k}$, $\Theta_S$ $=$ $\Theta_{S_1}$ $\cdots$ $\Theta_{S_m}$).

The next problem is to find an additional condition on the primary renormalization maps,
which would imply Eq.~(\ref{er6}).

Nevertheless, it is interesting to study the renormalization group even without
the unitarity condition.
In fact, in the Euclidean case the counterpart of the unitarity is just the fact
that the renormalization maps transform real functions to real distributions,
which is thus satisfied by definition.
Let us also point out that even if we construct time--ordered products, which do not obey
the unitarity then there is a simple, purely algebraic way to pass to a new
system of time--ordered products that satisfy all $(T_0)$--$(T_6)$ (\cite{EG}).

\Section{Action Ward identity}{se6.2}

This is the condition
\beq\label{AWI}
T\bigl(\di_{x_k^{\mu}}\Theta\bigr) \, = \,
\di_{x_k^{\mu}} T\bigl(\Theta\bigr)
\eeq
($\Theta \in \FSP_{\INFT \N}$, $k \in \N$, $\mu=0,\dots,D-1$).
In fact, since we are working only with translation invariant distributions
we have the translation invariance~of~$T$:
\beq\label{TRI}
\mathop{\sum}\limits_{k \, \in \, \N} \,
T\bigl(\di_{x_k^{\mu}}\Theta\bigr) \, = \,
\mathop{\sum}\limits_{k \, \in \, \N} \,
\di_{x_k^{\mu}} T\bigl(\Theta\bigr)
\,.
\eeq
Condition (\ref{AWI}) has been established in the off-shell (functional) approach
(\cite{DF})
but it has simple obstructions in the operator formalism.
For instance, for a free scalar field of mass $m$
the field equations together with the
causal Wick expansion
implies that
$$
\bigl(\Box_{\x_1} - m^2\bigr) \, T\bigl(\varphi (\x_1) \varphi(\x_2)\bigr)
= \delta(\x_1-\x_2) \neq
T\bigl(\bigl(\Box_{\x_1} - m^2\bigr) \varphi  (\x_1) \varphi(\x_2)\bigr) = 0 \,.
$$

A sufficient condition for Eq.~(\ref{AWI}) would be to require
a similar additional condition on the renormalization map $\RMA$:
\beq\label{RAWI}
\RMA\bigl(\di_{x_k^{\mu}}G\bigr) \, = \,
\di_{x_k^{\mu}} \RMA\bigl(G\bigr)
\eeq
($G \in \PA_{\N}$, $k \in \N$, $\mu=0,\dots,D-1$).
But now this contradicts to the requirement $(r3)$
that is also equivalent to the identities
\beq\label{RX}
\RMA\bigl(x_k^{\mu} \, G\bigr) \, = \,
x_k^{\mu} \, \RMA\bigl(G\bigr)
\,.
\eeq
The reason is that the two conditions, (\ref{RAWI}) and (\ref{RX}),
would imply that the renormalization map commute with the action
of all linear partial differential operators, which is not possible.
Of course,
one can ask why not to use
condition (\ref{RX}) instead of (\ref{RAWI})?
The proof of Lemma 2.7 in \cite{N1} shows
that in the case of (\ref{RAWI})
the recursion there goes in the wrong direction
(see also Remark~\ref{Rm}).
And in fact, there are again simple obstructions for Eq.~(\ref{RAWI}):
the function $(\x^2)^{-\frac{D}{2}+1}$
(on $D$--dimensional Minkowski space)
has a unique extension\footnote{%
since its scaling degree is less than $D$}
$\RMA \Bigl((\x^2)^{-\frac{D}{2}+1}\Bigr)$ and this is the Green function for
the D'Alembert operator;
hence,
$$
\Box_{\x} \, \RMA \Bigl((\x^2)^{-\frac{D}{2}+1}\Bigr) \, = \, \delta (\x) \, \neq \,
\RMA \Bigl(\Box_{\x} \, (\x^2)^{-\frac{D}{2}+1}\Bigr) \, = \, 0 \,.
$$

\Section{Renormalizing perturbative Euclidean field theory}{se6.3}

The Euclidean version of our ``theory of renormalization maps''
is easier but to formulate perturbative Euclidean QFT in the spirit
of algebraic perturbative QFT on Minkowski space is more difficult.
The problem is what would replace the time--ordered products in the Euclidean case.
Of course, we can use again the formula
$T\bigl(\Theta_S\bigr)$ $=$ $\RMA \bigl(\GH_S\bigr)$ $N\bigl(\Theta_S\bigr)$
(Eq.~(\ref{WTE1})) for a definition of the map $T$.
The problem then is that in the Euclidean case there are no
pointwise
Wick products!
In other words, $N\bigl(\Theta_S\bigr)$ do not generally exist as
densely defined operators on the Euclidean Fock space.
They can be only represented as quadratic forms.
For instance, one may see this problem even for Wick squares:
if $\varphi_E (\x)$ is the free Euclidean field of mass $m$ can we define
$\NPR{\varphi_E^2} (\x)$ as an operator on the Euclidean Fock space?
If so,
then
$$
\lvac \NPR{\varphi_E^2} (\x) \NPR{\varphi_E^2} (\y) \rvac \, = \, G_E (\x-\y)^2 \,.
$$
But in contrast to the Minkowski case $G_E (\x-\y)^2$ does not exists
(while the square or product of any Wightman functions
do exit).

A possible
way-out
is to work only off-shell (\cite{S1,K}) and hope that we can construct a positive functional
on the algebra of renormalized Euclidean time--ordered products.
However, it is unlikely that the latter is possible.
The reason is that such a positive functional will extend the vacuum functional on the initial
Euclidean field algebra (which is in fact, affiliated with a von Neumann commutative algebra)
and hence, it will be possible to represent the time--ordered products
on the Euclidean space, which already exist as quadratic forms there, also as operators.

\medskip

\noindent
\textbf{Acknowledgments.}
I am grateful to Professor R. Stora for his critical remarks to my paper \cite{N1},
which stimulated the present work.
I am also grateful to Professor I. Todorov for his comments.
This work was partially supported by the French--Bulgarian project Rila
under the contract Egide -- Rila N112 and by Bulgarian NSF grant DO 02--257.

\medskip

\newcommand{\Bibitem}[1]{\bibitem[#1]{#1}}
\newcommand{\BIbitem}[2]{\bibitem[#1]{#2}}
\addtocontents{toc}{\protect\vspace{-8pt}}
\addtocontents{toc}{\contentsline {section}{References}{\arabic{page}}}

\end{document}